# How well does *I3* perform for impact measurement compared to other bibliometric indicators?[1]

# The convergent validity of several (field-normalized) indicators

Lutz Bornmann*, Alexander Tekles,*[+] & Loet Leydesdorff[$]

*Administrative Headquarters of the Max Planck Society

Division for Science and Innovation Studies

Hofgartenstr. 8

80539 Munich, Germany.

Email: bornmann@gv.mpg.de; alexander.tekles.extern@gv.mpg.de

[+]Ludwig-Maximilians-University Munich

Department of Sociology

Konradstr. 6

80801 Munich, Germany.

[$]University of Amsterdam

Amsterdam School of Communication Research (ASCoR)

PO Box 15793,

1001 NG Amsterdam, The Netherlands.

Email: loet@leydesdorff.net

---

[1] This manuscript is an extended version of a document which was submitted to the 17th International Society of Scientometrics and Informetrics Conference (see https://www.issi2019.org)

**Abstract**

Recently, the integrated impact indicator (*I3*) indicator was introduced where citations are weighted in accordance with the percentile rank class of each publication in a set of publications. *I3* can also be used as a field-normalized indicator. Field-normalization is common practice in bibliometrics, especially when institutions and countries are compared. Publication and citation practices are so different among fields that citation impact is normalized for cross-field comparisons. In this study, we test the ability of the indicator to discriminate between quality levels of papers as defined by Faculty members at F1000Prime. F1000Prime is a post-publication peer review system for assessing papers in the biomedical area. Thus, we test the convergent validity of *I3* (in this study, we test *I3*/*N* – the size-independent variant of *I3* where *I3* is divided by the number of papers) using assessments by peers as baseline and compare its validity with several other (field-normalized) indicators: the mean-normalized citation score (MNCS), relative-citation ratio (RCR), citation score normalized by cited references (CSNCR), characteristic scores and scales (CSS), source-normalized citation score (SNCS), citation percentile, and proportion of papers which belong to the *x*% most frequently cited papers ($PP_{top\ x\%}$). The results show that the $PP_{top\ 1\%}$ indicator discriminates best among different quality levels. *I3* performs similar as (slightly better than) most of the other field-normalized indicators. Thus, the results point out that the indicator could be a valuable alternative to other indicators in bibliometrics.

**Key words**

bibliometrics, *I3*, field normalization, citation analysis, convergent validity

# 1 Introduction

In the application of citation analysis in research evaluation, one may need to compare the citation impact of publications from different fields. Different from using raw citation counts from the Web of Science (WoS, Clarivate Analytics) or Scopus (Elsevier) databases, professional bibliometricians have knowledge of differences in publication and citation cultures among fields of science (e.g., concerning the speed and frequency of citations) and use methods to assess the citation impact of focal papers against the impact of all other papers *in the same field and publication year* (McAllister, Narin, & Corrigan, 1983; Narin, 1981; Wang, Song, & Barabási, 2013). Field delineation, however, is not an easy task (e.g., Klavans & Boyack, 2017; Leydesdorff, 2006).

Various indicators (approaches) have been introduced in bibliometrics since the early 1980s to construct field-normalized scores. According to Waltman (2016) "the idea of these indicators is to correct as much as possible for the effect of variables that one does not want to influence the outcomes of a citation analysis, such as the field, the year, and the document type of a publication" (p. 375). The necessity to normalize citation impact for cross-field comparisons is also one of the ten principles for research evaluation formulated in the Leiden Manifesto (Hicks, Wouters, Waltman, de Rijcke, & Rafols, 2015).

Leydesdorff and Bornmann (2011b) introduced the integrated impact indicator (*I3*) where citations are weighted in accordance with the percentile rank class of each publication in a set of publications (e.g., published by a researcher or research group). Percentiles are *a priori* field-normalized: one can compare the top-1% for different reference sets. Although several publications appearing afterwards have dealt with the indicator (Leydesdorff & Bornmann, 2012; Rousseau, 2012; Wagner & Leydesdorff, 2012; Ye, Bornmann, & Leydesdorff, 2017), a comparison with other (field-normalized) indicators has not yet been done. In this study, we undertake this comparison by investigating the convergent validity of the indicator. In psychometrics, convergent validity tests whether measurements which are assumed to be related (here: assessments by peers and citation impacts) are actually related: we are interested in the question of how *I3* discriminates between papers having received different quality scores by peers compared to various other indicators. We received a dataset from F1000Prime (see https://f1000.com/prime) including the bibliographic information of papers published in the biomedical area and their quality scores by peers. We use these scores as a benchmark for testing the indicators (Garfield, 1979).

## 2 Normalization of citation impact in bibliometrics

In this section, we discuss the various field-normalized indicators which are used for the comparison with the *I3* indicator: mean-normalized citation score (MNCS), relative-citation ratio (RCR), citation score normalized by cited references (CSNCR), characteristic scores and scales (CSS), source normalized citation score (SNCS), citation percentile, and proportion of papers which belong to the *x*% most frequently cited papers ($PP_{top\ x\%}$). More comprehensive overviews of methods for normalizing citations can be found in Mingers and Leydesdorff (2015), Waltman (2016), and Bornmann (in press). *I3* is explained in section 2.7 after all other indicators have been explained, since the *I3* variant used in this study is based on other field-normalizing approaches.

One can distinguish between field-normalization and statistical normalization: each indicator assumes some form of reference sets (field-normalization) and some form of comparison-strategy (statistical normalization). The indicators compared in this study vary with respect to both these aspects: different reference sets (e.g., papers published in the same subject category or co-cited papers) and different strategies to compare the focal papers to these reference sets (e.g., comparing values in relation to the mean or generating non-parametric percentiles for the comparison). Most of the variance among the indicators selected for this study are a result of the statistical normalization. However, there is always already (at least some) variance among the indicators with respect to the field categorizations (i.e., the indicators have not been calculated by using one single categorization scheme). Most of the indicators in this study have been calculated based on WoS subject categories (WCs). However, RCR and the citing side indicators are not relying on these categories, but on co-cited papers and papers published in the same journal or paper.

The use of WCs for field-normalization has been criticized as unprecise in terms of its analytical basis. WCs are attributed to journals (and not to individual papers) and journals are not homogeneous in terms of the disciplines of papers published in them (Leydesdorff & Bornmann, 2016). Although other field-categorisation schemes have been proposed for the normalization of citation impact such as algorithmically constructed classification systems (Ruiz-Castillo & Waltman, 2015) or expert-based field categorisations (Bornmann, Marx, & Barth, 2013) "the WoS journal subject categories are the most commonly used field classification system for normalisation purposes" (Wouters et al., 2015, p. 18).

All indicators considered here with the exception of the RCR are available at the paper level in an in-house database of the Max Planck Society which is based on the Web of Science (WoS, Clarivate Analytics). We retrieved the additional RCR scores in a two-step process. First, the DOIs are used to automatically query the papers' Pubmed IDs using the web form available under https://icite.od.nih.gov/analysis. Each of these requests returns an HTML document containing the Pubmed ID of the corresponding papers. Second, the Pubmed IDs were extracted from the HTML documents and used for requesting the papers' RCR scores via the iCite API at https://icite.od.nih.gov/api.

## 2.1 Mean-normalized citation score (MNCS)

Based on early proposals by Schubert and Braun (1986), Moed, Burger, Frankfort, and van Raan (1985) used field-normalization based on the WCs in the so-called "crown indicator" of the Leiden Centre for Science and Technology Studies (CWTS). Opthof and

Leydesdorff (2010) note that the statistical normalization in the definition of the crown indicator

$$(\sum observed / \sum expected) \qquad (1)$$

was statistically erroneous (see Lundberg, 2007). Given the order of operations, one should first multiply and divide and only thereafter sum and subtract (Gingras & Larivière, 2011; Leydesdorff & Opthof, 2018). The normalization can then be formulated as follows:

$$\frac{1}{n} \sum \frac{observed}{expected} \qquad (2)$$

In a response, Waltman, van Eck, van Leeuwen, Visser, and van Raan (2011) proposed to use this "mean-normalized citation score" (MNCS) with field-normalization by defining the mean in the denominator of each paper in terms of the WCs attributed to the respective journals. MNCS is currently a frequently used field-normalized indicator (Purkayasthaa, Palmaroa, Falk-Krzesinskib, & Baas, 2018). It is calculated by dividing the citations of a paper in question by the average citation rate of the papers that were published in the same subject category (and publication year).

Two normalizations are thus involved: (1) normalization relative to the mean and (2) normalization in terms of WCs. MNCS, however, can also be used with classification schemes other than WCs. The first assumption that the mean of the citation rate of the papers in the sample can be considered as an expected value, is not valid. The citation distributions are always skewed and thus non-normal. (The Central Limit Theorem is only valid for much larger samples.) At the time (2011), we proposed the use of percentile classes instead (Bornmann & Mutz, 2011; Leydesdorff, Bornmann, Mutz, & Opthof, 2011).

A further complication arises when a paper is published in a journal that belongs to more than a single subject category. MNCS can then be calculated with reference to different sets, e.g., by using "fractional counting" (Smolinsky, 2016; Waltman et al., 2011). In this study, the average is calculated over the MNCSs in the case of multiple categories. The impact of different publication sets can then be compared by using the mean of the MNCSs.

## 2.2 Relative-citation ratio (RCR)

Hutchins, Yuan, Anderson, and Santangelo (2016) proposed the Relative Citation Rate (RCR) as a new field-normalized impact indicator. The indicator is similarly designed as MNCS: it is a quotient of the focal paper's citation counts and the expected number of citations in the reference set. The difference of the RCR from the MNCS is that the expected value (respectively the reference sets) is based on co-citations: the papers co-cited with the focal paper are considered to represent a more precise reference set at the paper level than WCs which are attributed at the journal level. In bibliometrics, co-citations are frequently used similarity measures which are based on citation relations. An overview of research on the RCR can be found in Lindner, Torralba, and Khan (2018).

## 2.3 Citation score normalized by cited references (CSNCR)

Bornmann and Haunschild (2016) introduced the field-normalized indicator "citation score normalized by cited references" (CSNCR) which is closely related to the MNCS. The indicator is rooted in early suggestions by Garfield (1979) that "the most accurate measure of citation potential is the average number of references per paper published in a given field". The CSNCR is defined as follows: the citations of a focal paper are divided by the mean number of cited references in a subject category. The theoretical analysis of the CSNCR by Bornmann and Haunschild (2016) demonstrated that the indicator has the properties of consistency and homogeneous normalization. The authors' empirical comparison of the CSNCR with other field-normalized indicators revealed that it is as suitable as other field-normalized indicators to normalize citations.

## 2.4 Characteristic scores and scales (CSS)

The characteristic scores and scales (CSS) method by Glänzel and Schubert (1988) for normalizing citation data is one of the earliest proposed field-normalization approaches. The CSS method classifies the publications in reference sets (subject categories) as follows: "characteristic scores are obtained from iteratively truncating a distribution according to conditional mean values from the low end up to the high end. In particular, the scores $b_k$ (k > 0) are obtained from iteratively truncating samples at their mean value and recalculating the mean of the truncated sample until the procedure is stopped or no new scores are obtained" (Glänzel, 2013, p. 111). In many studies based on this method, four impact classes are used to group the papers in reference sets (see Glänzel, Thijs, & Debackere, 2014):

1. poorly cited (papers with less citations than $b_1$),
2. fairly cited (papers with citations above $b_1$ but less citations than $b_2$),
3. remarkably cited (papers with citations above $b_2$ but less citations than $b_3$), and
4. outstandingly cited (papers with citations of at least $b_4$).

In the MPG in-house database, all papers in each reference set published since 1980 are classified following the CSS method.

## 2.5 Citing-side normalization of citation impact

Citations are attributed to papers on the cited side by the indicators mentioned above. Zitt and Small (2008) first introduced the idea of normalizing citation impact on the citing-side. The authors proposed a modification of the journal impact factor (JIF) by fractional citation weighting. Citing-side normalization is also named source normalization, fractional citation weighting, fractional counting of citations, or a priori normalization (Waltman & van Eck, 2013a). The method cannot only be used for journals as initially proposed by Zitt and Small (2008) but also for any other publication sets (Moed, 2010). Citing-side normalization considers the environment of a given citation (Leydesdorff & Bornmann, 2011a; Leydesdorff, Radicchi, Bornmann, Castellano, & de Nooy, 2013): the citation is weighted depending on its environment. A citation from a subject category with papers containing long reference lists (e.g., bio-medicine) receives a lower weighting than a citation from a subject category with on average only few citations.

For citing-side normalization, the number of references of the citing paper is usually used to weight a specific citation (Waltman & van Eck, 2013b). The assumption is that this number of references reflects the typical number in the field (subject categories) of the citing paper. However, this assumption cannot always be made. For this reason, an average number of references is calculated (and used as weighting factor) which includes other papers appearing in a journal alongside the citing paper. In this study, we consider three variants of citing-side normalization, which are explained by Waltman and van Eck (2013b) in more detail.

**Variant 1:**

$$SNCS1 = \sum_{i=1}^{c} \frac{1}{a_i} \qquad (5)$$

The first variant is the SNCS1 (source normalized citation score) indicator. In the formula, $a_i$ is the average number of *linked* references in those papers which appeared in the same journal and in the same publication year as the citing paper $i$. "Linked references" are references to papers in journals covered by the WoS. The reduction to linked references (instead of using all references) is intended to prevent that subject categories of citing publications not indexed in WoS are disadvantaged (see Marx & Bornmann, 2015). For the calculation of the average number of linked references, only those from specific reference publication years are considered. The number of the considered publication years in the references is defined as the citation window (for the cited publications). For example, if the citation window of the cited paper (published in 2008) is four years (2008 to 2011), each citation of this paper is divided by the average number of linked references from the previous four years. Analogously, a citation from 2010 would be divided by the average number of linked references from 2007 to 2010. The focus on recent publication years is intended to prevent subject categories in which older literature plays a significant role, to be disadvantaged in the normalization (Waltman & van Eck, 2013b).

**Variant 2:**

$$SNCS2 = \sum_{i=1}^{c} \frac{1}{r_i} \qquad (6)$$

For the second variant SNCS2 each citation of a paper is divided by the number of linked references in the citing publication $r_i$. The difference to SNCS1 is that SNCS2 focusses on the linked references in the citing paper and not the journal of the citing paper. The selection of the reference publication years is done analogously to the SNCS1.

**Variant 3:**

$$SNCS3 = \sum_{i=1}^{c} \frac{1}{p_i r_i} \qquad (7)$$

SNCS3 combines SNCS1 and SNCS2. $r_i$ is defined as in SNCS2. $p_i$ is the paper share containing at least one linked reference among the papers in the same journal and publication year as the citing publication $i$. The selection of the reference publication years follows the same procedure as for the SNCS1 and SNC2.

## 2.6 Percentile-based indicators

2.6.1 Citation impact percentiles

The distribution of citation data is usually very skewed with only a few papers being highly-cited (Seglen, 1992). Since the arithmetic mean is not appropriate as a measure of the central tendency in a skewed distribution, citation impact percentiles have been introduced as an alternative to approaches based on the averages of citations. The citation impact percentile of a specific paper indicates the share of other papers in the reference set which have received fewer citations. For example, a citation impact percentile of 80 indicates that 80% of the papers in the reference set have received fewer citations.

Citation impact percentiles from different reference sets are directly comparable with one another; no further field-normalization is needed. Suppose the citation impact of two papers have been normalized based on different reference sets and both papers have a percentile of 70. The identical percentile indicates that both papers have – compared with the other papers in the corresponding reference sets – achieved the same citation impact. Even though both papers may have different times cited values in the WoS database, the relative citation impacts are the same.

Citation impact percentiles can be calculated with various procedures (see the overview in Bornmann, Leydesdorff, & Mutz, 2013). In the current study, two approaches were used which are frequently applied in evaluative bibliometrics. For both approaches, all papers in the reference sets are ranked in decreasing or increasing order by their citation counts ($i$), and the number of publications in the reference set ($n$) is determined in the first step. For the product InCites (a customised, web-based research evaluation tool based on bibliometric data from WoS), Clarivate Analytics calculates the percentiles by using (basically) the formula ($[i/n] * 100$). This inversed ranking will be named as "InCites percentiles" in the following. However, the use of this formula may lead to a mean percentile of a reference set unequal to 50 (the median). The formula ($[(i - 0.5)/n] * 100$) (Hazen, 1914) does not suffer this disadvantage. We will use the abbreviation "Hazen percentile" for these percentiles in the following. Furthermore, the papers are sorted in increasing impact order for InCites percentiles, but in decreasing order for Hazen percentiles; we invert the InCites percentiles in this study by subtracting the values from 100.

2.6.2  Proportion of papers belonging to the top-$x$%

Citation percentiles can be directly used for impact measurements. However, it is also very common in bibliometrics to focus on certain percentile classes (Bornmann, 2014). In this study, we include three indicators focusing on three classes: $PP_{top-50\%}$, $PP_{top-10\%}$, and $PP_{top-1\%}$. The indicators reveal the proportion of papers published by a unit which belong to the $x$% most frequently cited papers. The results of Tahamtan and Bornmann (2018) show that the $PP_{top-x\%}$ indicators – especially the $PP_{top-10\%}$ indicator – are one of the earliest used field-normalized indicators in scientometrics which were introduced by Narin (1981). In this study, we used $PP_{top-x\%}$ indicators which have been calculated based on two fractional counting approaches.

Papers may be equal in the rankings, if the papers are sorted by citations and more than one paper have the same citation counts. These ties in citations lead to the problem of exactly assigning the papers to the top-$x$% class or the corresponding bottom-$x$% class. To solve this problem we use an approach introduced by Waltman and Schreiber (2013). They propose to fractionally assign the papers at the top-$x$% threshold to the top- and bottom-$x$% – in dependence of the number of papers with the same number of citations at the threshold.

The second fractional counting approach used for the indicators concerns the multiple assignment of journals to subject categories. We use the fractional counting approach by Waltman et al. (2011) to calculate the PP$_{top-x\%}$ indicators across multiple subject categories.

**2.7  *I3* indicator**

One of the newest (*a priori* field-normalized) indicators is the integrated impact indicator (*I3*) which is also percentile-based. It was defined as a non-parametric alternative in response to the above mentioned discussions about statistical normalization of the CWTS "crown-indicator" (Leydesdorff et al., 2011). Bornmann (2010) and Bornmann and Mutz (2011) proposed to use the weighted number of papers of units (e.g., journals or institutions) belonging to certain percentile impact classes for performance measurements. The further elaboration into *I3*, the integrated impact indicator, combines these proposals in a unified scheme (Leydesdorff & Bornmann, 2011b; Leydesdorff & Bornmann, 2012; Rousseau, 2012; Wagner & Leydesdorff, 2012).

In the most recent development, Leydesdorff, Bornmann, and Adams (2018) propose to use four percentile classes (top-1%, top-10%, top-50%, and bottom-50%) as weighting scheme for *I3*. They argued that a paper in the top-1% class can be valued ten times more than a paper in the top-10% class. It follows that a top-1% paper weights 100 times a paper at the bottom. It is an advantage of this scheme that it appreciates the highly-skewed nature of citation data by using a logarithmic scale. It follows that papers in the top-50% are weighted with two and bottom-50% with one. The resulting indicator correlates above .9 with the numbers of both publications and citations in empirical cases.

Table 1 shows an example for the calculation of *I3* which is based on publication and citation data for the journal *PLOS One*. The numbers in the third column are the numbers of papers in the different top-*x*% classes (fractionally counted, see above). Since the paper numbers in the higher top-*x*% classes are subsets of the numbers in the lower classes, the numbers in the percentile classes have been corrected correspondingly to avoid double counting of papers. The corrected numbers in the classes are multiplied by the weights. As the last column in Table 1 reveals, the weighted numbers of papers in the distinct classes result in *I3* which is 53,570.256. This field-normalized number can be compared with *I3* values for any other journal (or other document set) for performance measurements considering both the output- and impact dimension.

Table 1. Publication and citation data of *PLOS One* as an example for calculating *I3*. Source: Leydesdorff et al. (2018)

| Classes | Percentile threshold | Number of papers | Weight | *I3* |
|---|---|---|---|---|
| top-1% | 99-100 | 14 | x 100 = | 1400 |
| top-10% | 90-98 | 912.821 | x 10 = | 9128.21 |
| top-50% | 50-89 | 13,926.867 | x 2 = | 27853.734 |
| bottom-50% | 0-49 | 15,188.312 | x 1 = | 15188.312 |
| Total | | 30,042 | | 53,570.256 |

Empirically derived *I3* values can be compared with theoretically possible values. The minimal possible *I3* of *PLOS One* 2014 is 30,042. In this case, all papers would belong to the bottom-50% papers which are weighted with 1 (0 * 100 + 0 * 10 + 0 * 2 + 30,042 * 1 = 30,042). In contrast, the maximal possible *I3* is 3,004,200 (30,042 * 100 + 0 * 10 + 0 * 2 + 0 * 1). The maximum can be reached with all papers in the top-1% most frequently cited percentile class. With *I3* = 53,570.256, the journal reaches 1.78% of the maximum.

# 3    Methods

## 3.1    Peer ratings provided by F1000Prime

F1000Prime is a post-publication peer review system of papers published in medical and biological journals. The service started with F1000 Biology in 2002; F1000 Medicine followed in 2006. Both services were merged in 2009 to the current F1000Prime database.

Papers which are included in the F1000Prime database are selected by a peer-nominated global "Faculty". These are leading scientists and clinicians who assess the papers and explain their importance. F1000Prime covers a restricted set of papers published in medical and biological journals (Kreiman & Maunsell, 2011; Wouters & Costas, 2012).

The Faculty includes more than 5,000 experts worldwide. Faculty members can choose and assess any paper of interest. Although many papers published in popular and reputable journals (e.g., *Nature* and *Science*) are evaluated by the members, most of the papers have been published in specialised or less well-known journals (Wouters & Costas, 2012). "Less than 18 months since Faculty of 1000 was launched, the reaction from scientists has been such that two-thirds of top institutions worldwide already subscribe, and it was the recipient of the Association of Learned and Professional Society Publishers (ALPSP) award for Publishing Innovation in 2002 (http://www.alpsp.org/about.htm)" (Wets, Weedon, & Velterop, 2003, p. 249).

The selected papers for F1000Prime are rated by the Faculty members as "good", "very good", or "exceptional" which are set to the scores of 1, 2, or 3, respectively. Since many papers are assessed not only by one Faculty member but by several, we calculated the sum of the scores for this study. This accords to the F1000Prime practice to use the individual scores for calculating the total score for each paper (which are used then to rank the papers in the disciplines). The assessments in the F1000 database can be used either by scientists for receiving pointers to relevant papers in their areas, but also as a database for research evaluation purposes. According to Wouters and Costas (2012) "the data and indicators provided by F1000 are without doubt rich and valuable, and the tool has a strong potential for research evaluation, being in fact a good complement to alternative metrics for research assessments at different levels (papers, individuals, journals, etc.)" (p. 14).

## 3.2 Used datasets

In 2018, F1000 provided one of the authors with data on recommendations made by the Faculty members and the bibliographic information for the corresponding papers in their system ($n$=51,461 papers). We matched the papers with the papers in our WoS in-house database (of the Max Planck Society) using the DOI. We restricted the set to papers with the document types "article" and "review". In the statistical analyses, we included not only the field-normalized indicators explained in section 2 (with a citation window between publication year and the end of 2017), but also citation counts (1) for a three-years citation window and (2) for the period between publication year and the end of 2017. We included only matched F1000Prime papers into the study until 2015 to ensure a minimum citation window of three years (Glänzel & Schöpflin, 1995). Since the indicators have different numbers of missing values, only papers have been considered with no missing value across all indicators. These restrictions lead to a total number of 28,063 papers for the statistical analysis published between 2000 and 2015 (see Table 2). Most of the reduction is due to the necessity of using a minimum citation window.

Table 2. Number of papers in the dataset across publication years

| Publication year | Number of papers | Number of papers (in percent) |
|---|---|---|
| 2000 | 2 | 0.01 |
| 2001 | 2 | 0.01 |
| 2002 | 38 | 0.14 |
| 2003 | 44 | 0.16 |
| 2004 | 88 | 0.31 |
| 2005 | 163 | 0.58 |
| 2006 | 204 | 0.73 |
| 2007 | 224 | 0.8 |
| 2008 | 300 | 1.07 |
| 2009 | 414 | 1.48 |
| 2010 | 617 | 2.2 |
| 2011 | 901 | 3.21 |
| 2012 | 2589 | 9.23 |
| 2013 | 8987 | 32.02 |
| 2014 | 7177 | 25.57 |
| 2015 | 6313 | 22.5 |
| Total | 28,063 | 100 |

Since this study is based on papers published in medical and biological journals, one can doubt whether the dataset is useful for comparing field-normalized indicators. The dataset might be too focussed on narrow research areas to study the ability of different methods for cross-disciplinary normalization. Thus, we took a look at the WCs to which the papers (or journals) in our dataset have been assigned to. Is there really that much difference in the WCs that the dataset can be used in this study? We focused on WCs, since we used them for field-normalization in this study. As the results show, each F1000Prime paper was assigned to up to six different WCs. In WoS, around 250 WCs exist; at least one paper in our dataset was assigned to one of 157 WCs.

Table 3. Minimum and maximum of the mean number of citations, authors, and cited references in 157 WCs including F1000Prime papers. Furthermore, the minimum and maximum number of papers in the WCs are shown.

| Statistics | Mean number of citations | Mean number of citations (three-year citation window) | Mean number of authors | Mean number of cited references | Number of papers |
|---|---|---|---|---|---|
| Minimum | 4.4 | 3.0 | 1.0 | 16.6 | 1 |
| Maximum | 1581.0 | 125.3 | 18.0 | 83.4 | 5466 |

Subject-specific differences in publication and citation cultures are usually revealed by differences in the mean number of citations, authors, and cited references. Table 3 shows the minimum and maximum of the mean number of citations, authors, and cited references in the 157 WCs (in addition to the minimum and maximum of the number of papers). The number of papers in the WCs differs between 1 and 5466. As the results in Table 3 point out, the F1000Prime dataset is concerned by larger differences in the mean numbers of citations, authors, and cited references. Since these results point to larger subject-specific heterogeneity in the dataset, it might be reasonable to use the dataset for studying the validity of different methods for cross-disciplinary normalization.

**3.3  Software used**

The statistical software package Stata 15.1 (http://www.stata.com/) was used for this study (StataCorp., 2017).

# 4  Results

We included 14 (field-normalized) indicators in this study for comparing them with *I*3. As the explanations of the indicators in section 2 show, the indicators have different levels of measurement. For example, the CSS indicator is a variable with ordinal scale; the SNCS indicators are variables with cardinal scale. However, in the scale assignments to the indicators it should be born in mind that the possibility of multi-assignment of papers to WCs are considered in the calculation. For example, in case of percentiles, average values have been calculated. The $P_{top\ x\%}$ indicators are not binary variables, because we used the fractional counting approach to calculate the indicator values across multiple WCs (see above).

Table 4. Spearman rank order correlations between field-normalized indicators (besides *I*3)

| Indicators | $P_{top\ 50\%}$ | $P_{top\ 10\%}$ | $P_{top\ 1\%}$ | Number of citations (until 2017) | Number of citations (three year citation window) | MNCS | CSNCR | CSS | SNCS1 | SNCS2 | SNCS3 | Hazen percentiles | Incites percentiles | RCR |
|---|---|---|---|---|---|---|---|---|---|---|---|---|---|---|
| $P_{top\ 50\%}$ | 1.00 | | | | | | | | | | | | | |
| $P_{top\ 10\%}$ | 0.38 | 1.00 | | | | | | | | | | | | |
| $P_{top\ 1\%}$ | 0.14 | 0.38 | 1.00 | | | | | | | | | | | |
| Number of citations (until 2017) | 0.53 | 0.79 | 0.51 | 1.00 | | | | | | | | | | |
| Number of citations (three year citation window) | 0.53 | 0.82 | 0.52 | 0.96 | 1.00 | | | | | | | | | |
| MNCS | 0.55 | 0.87 | 0.55 | 0.90 | 0.93 | 1.00 | | | | | | | | |
| CSNCR | 0.54 | 0.83 | 0.52 | 0.90 | 0.91 | 0.96 | 1.00 | | | | | | | |
| CSS | 0.34 | 0.74 | 0.52 | 0.85 | 0.80 | 0.82 | 0.80 | 1.00 | | | | | | |
| SNCS1 | 0.54 | 0.85 | 0.54 | 0.92 | 0.94 | 0.94 | 0.94 | 0.80 | 1.00 | | | | | |
| SNCS2 | 0.51 | 0.81 | 0.53 | 0.89 | 0.90 | 0.90 | 0.90 | 0.77 | 0.96 | 1.00 | | | | |
| SNCS3 | 0.51 | 0.80 | 0.53 | 0.88 | 0.89 | 0.90 | 0.90 | 0.76 | 0.96 | 1.00 | 1.00 | | | |
| Hazen percentiles | 0.55 | 0.87 | 0.55 | 0.90 | 0.93 | 1.00 | 0.96 | 0.82 | 0.94 | 0.90 | 0.89 | 1.00 | | |
| Incites percentiles | 0.55 | 0.88 | 0.57 | 0.90 | 0.93 | 0.98 | 0.93 | 0.81 | 0.95 | 0.91 | 0.91 | 0.98 | 1.00 | |
| RCR | 0.51 | 0.82 | 0.53 | 0.87 | 0.90 | 0.92 | 0.90 | 0.77 | 0.94 | 0.90 | 0.90 | 0.91 | 0.92 | 1.00 |

To have a first overview of the different (field-normalized) indicators, we calculated Spearman rank order correlations (see Table 4). As the correlation coefficients in the table reveal, most of the coefficients are on a large or (much) larger than typical level (following the guidelines by Cohen, 1988, to interpret correlation coefficients). This is also the case for the correlations between normalized and non-normalized indicators (i.e., number of citations). The results in Table 4 might be interpreted as first hints that the differences between the indicators in measuring citation impact (field-normalized) are not very large. However, we could not include *I3* in the correlation analysis, since *I3* can only be used on the aggregated (group) level.

Since *I3* can be used as a field-normalized indicator, we are interested in this study in how it discriminates between papers rated differently by Faculty members compared to other (field-normalized) indicators (see section 2). In other words, we are interested in its convergent validity: does the indicator discriminate worse, equal to, or better than the other indicators between the different quality levels and is thus more convergently valid to the assessment by peers than the other indicators? *I3* differs from the other indicators by being calculated on the aggregated, and not on the single paper level. Thus, we need groups of papers for the comparison of *I3* with other indicators.

The CSS method which we explained in section 2.4 cannot only be used to field-normalize single papers, but to group any paper set with metrics (see, e.g., Bornmann & Glänzel, 2018). Using the CSS method to group the papers in four classes – based on the sum of the F1000Prime scores per paper – we found 1396 papers (4.97%) in the class with the best scores (F1000 class 4, sum scores between 5 and 35), 3737 papers (13.32%) in the second best class (F1000 class 3, sum scores between 3 and 4), 10,334 papers (36.82%) in the next class (F1000 class 2, sum scores equal to 2), and 12,596 papers (44.88%) in the lowest class (F1000 class 1, sum scores equal to 1).

For the four groups, we calculated the arithmetic average of each indicator per group. The median would have been an alternative, but this statistic fails to properly differentiate between the groups because of ties in certain indicator values. For example, the $PP_{top\ x\%}$ indicators mostly consists of the values 0 and 1 which lead to corresponding indifferent median values for the classes. We decided not to use the sum, since the results are dependent on the sample size: the more papers in a group are, the better results can be expected.

Although *I3* was designed to reflect the output in addition to the impact dimension (as a sum score), the output dimension is not relevant for this validity study. The performance of the four F1000 classes is not dependent on the output dimension; only the impact of the single papers matters. In the usual evaluation of research groups or institutions, however, we are faced with a different situation in which both dimensions – publications and citations – are of equal interest for assessing performance.

In case of the *I3* indicator, we divided *I3* by the number of papers in a group and obtain *I3*/*N*. This has been proposed already by Leydesdorff et al. (2018) for the comparison of journal *I3* scores with the Journal Impact Factor (which is a *mean* citation rate). For the four F1000Prime quality groups, we received the following *I3*/*N* values: F1000 class 1 = 11.68, F1000 class 2 = 14.63, F1000 class 3 = 22.66, and F1000 class 4 = 39.03. The mean values point out that *I3* measures quality as expected: it discriminates validly between the four performance groups. However, does *I3/N* discriminate better between the groups than the other indicators (and is thus more convergently valid)? As the results in Table 5 show, all other indicators which we considered in this study are similarly able to discriminate between the four F1000 classes.

Table 5. Mean indicator scores for four F1000 classes (class 4 reflects the highest quality level)

| Indicator (mean value) | F1000 class | | | |
|---|---|---|---|---|
| | 1 ($n$=12,596) | 2 ($n$=10,334) | 3 ($n$=3737) | 4 ($n$=1396) |
| $PP_{top\ 50\%}$ | 0.87 | 0.91 | 0.96 | 1.00 |
| $PP_{top\ 10\%}$ | 0.41 | 0.50 | 0.68 | 0.89 |
| $PP_{top\ 1\%}$ | 0.07 | 0.10 | 0.17 | 0.33 |
| Number of citations (until 2017) | 54.03 | 58.77 | 93.92 | 157.69 |
| Number of citations (three year citation window) | 31.08 | 38.62 | 60.42 | 115.69 |
| MNCS | 3.18 | 3.68 | 5.60 | 9.92 |
| CSNCR | 4.19 | 4.79 | 7.59 | 13.68 |
| CSS | 0.59 | 0.64 | 0.89 | 1.24 |
| SNCS1 | 3.57 | 4.00 | 6.07 | 10.71 |
| SNCS2 | 3.14 | 3.53 | 5.28 | 9.18 |
| SNCS3 | 3.30 | 3.71 | 5.54 | 9.57 |
| Hazen percentiles | 78.43 | 82.92 | 89.25 | 95.93 |
| Incites percentiles | 78.27 | 82.76 | 89.27 | 95.84 |
| RCR | 3.72 | 4.15 | 6.42 | 11.54 |
| *I3* | 11.68 | 14.63 | 22.66 | 39.03 |

To compare the ability of the indicators to discriminate between the four F1000 classes, we calculated the so called "Average Annual Growth Rate (AAGR)" (instead of annual differences we have quality group differences in our study). The AAGR is the average increase in citation impact over the quality groups. It is computed by taking the arithmetic average of a series of growth rates. In the first step of calculating AAGR for each indicator, we determined the percentage growth for each group (except for F1000 class 1) which is the percentage growth (F1000 class $x$ / F1000 class $x – 1$) – 1. In the second step, the AAGR is calculated as the sum of each indicator's growth rate divided by the number of F1000 classes – 1. We also calculated the "Sum Annual Growth Rate (SAGR) for comparison with the AAGR which is a measure of the total increase in citation impact over the quality groups.

Table 6. AAGR and SAGR for the various indicators. The indicators are ordered by SAGR (and AAGR) in decreasing order. The column "Difference to previous SAGR" shows how much the SAGR of an indicator differs from its previous SAGR with the rank $x-1$.

| Indicator (mean value) | AAGR | SAGR | Rank | Difference to previous SAGR |
|---|---|---|---|---|
| $PP_{top\ 1\%}$ | 67.24 | 201.72 | 1 | |
| Number of citations (three year citation window) | 57.39 | 172.17 | 2 | -29.55 |
| CSNCR | 50.99 | 152.98 | 3 | -19.19 |
| *I3* | 50.79 | 152.38 | 4 | -0.60 |
| RCR | 48.70 | 146.10 | 5 | -6.28 |
| MNCS | 48.35 | 145.04 | 6 | -1.06 |
| SNCS1 | 46.67 | 140.01 | 7 | -5.03 |
| Number of citations (until 2017) | 45.50 | 136.49 | 8 | -3.52 |
| SNCS2 | 45.26 | 135.78 | 9 | -0.71 |
| SNCS3 | 44.80 | 134.41 | 10 | -1.37 |
| $PP_{top\ 10\%}$ | 29.52 | 88.57 | 11 | -45.84 |
| CSS | 29.16 | 87.48 | 12 | -1.09 |
| Incites percentiles | 6.99 | 20.97 | 13 | -66.51 |
| Hazen percentiles | 6.95 | 20.84 | 14 | -0.12 |
| $PP_{top\ 50\%}$ | 4.49 | 13.48 | 15 | -7.37 |

The results on the basis of AAGR and SAGR for the various indicators are shown in Table 6.[2] The indicators are sorted by SAGR (and AAGR) in decreasing order. The column "Difference to previous SAGR" reveals how much the SAGR of an indicator differs from the SAGR of the indicator with the rank $x-1$. Thus, the column indicates how much larger the scores in the better class are. The results in Table 6 point out that $PP_{top\ 1\%}$ discriminates best between the different quality classes. The indicator is followed by the number of citations (measured across a three year citation window). CSNCR is on the third position whereby *I3* has very similar AAGR and SAGR values as CSNCR.

As the "Difference to previous SAGR" column reveals, $PP_{top\ 1\%}$ discriminates much better than the second best positioned indicator number of citations (measured across a three year citation window) which performs itself much better than the CSNCR indicator. The indicators with the rank positions 3 to 10 are able to discriminate similarly between the four quality levels. The next larger performance difference are visible between $PP_{top\ 10\%}$ and SNCS3 (-45.84%) as well as between Incites percentiles and CSS (-66.51%).

---

[2] Normalized linear regression slopes reveal similar results. Thus, the results are stable independent on the used method.

## 5 Discussion: limitations and perspectives

The discussion about the normalization of citation impact has a long tradition in bibliometrics. Since publication and citation practices are very different among the various fields of science, citation numbers from different fields cannot be directly compared (Bornmann & Marx, 2015). The use of field-normalized indicators in research evaluation is one of the guiding principles in the Leiden Manifesto (Hicks et al., 2015). The same Manifesto advocates the use of percentiles for field normalization. In many evaluation contexts one uses field-normalized indicators (based on statistical normalization by the mean) for measuring citation impact instead of using the raw times-cited information from the WoS or Scopus databases. For example, field-normalized indicators are used in the popular Times Higher Education Rankings (see https://www.timeshighereducation.com/world-university-rankings).

Research on these indicators focused especially on the use of the arithmetic average of highly-skewed citation distributions. This poses a problem, for instance, for the use of MNCS and the way in which "research fields" are operationalized. Various categorization schemes can be used to define fields (e.g., schemes based on citation relations or subject categorizations from field-specific literature databases) and fields can be defined at different levels of aggregation (Wilsdon et al., 2015). Some research has been undertaken hitherto to identify field-normalized indicators using methods which normalize citation impact better than other indicators. According to the empirical results of Waltman and van Eck (2013a, 2013b), citing-side normalization has been shown more successful than cited-side normalization in field-normalizing citation impact. Purkayasthaa et al. (2018) reported the following results: "from the high correlations within our analyses of the two metrics across a range of research areas, we conclude that $RCR_{Scopus}$ and FWCI [field-weighted citation impact] can be used interchangeably to evaluate citation impact of an article or of larger entities such as universities". Bornmann and Leydesdorff (2013) and Bornmann and Marx (2015) used assessments from F1000Prime to compare the validity of different (field-normalized) citation impact indicators.

We included a range of (field-normalized) indicators in the current study to compare the newly proposed *I3* indicator with other indicators with respect to their convergent validity (using assessments by peers as a baseline; sometimes called the "golden standard" of peer review). We wanted to know whether *I3* is better able than other indicators in discriminating between different quality levels as defined by Faculty members working for F1000Prime. The indicators differ in terms of field-categorization (e.g., papers in the same WC or co-cited papers) and comparison-strategy (e.g., comparison of percentiles or focal papers with mean values). The investigation of the different indicators show smaller differences between different types of reference sets (field-categorization), but larger differences with respect to the comparison strategy (statistical normalization).

The results show that the $PP_{top\ 1\%}$ indicator discriminates best compared to the other indicators given the assumed baseline of F1000Prime. However, this result reflects the orientation of F1000Prime towards excellence in biomedicine which the $PP_{top\ 1\%}$ indicator targets more precisely than any of the other indicators. The second best indicator is the raw number of citations in the first three years after publication. Although this indicator is not field-normalized nor statistically normalized, it performs comparably well – perhaps because it focusses specifically on the period when most of the papers are selected by the Faculty members for inclusion in the F1000Prime database. The Faculty members might also consider the number of citations in their selection decisions and assessments of the papers.

Furthermore, the F1000Prime dataset is a relatively homogenous dataset with respect to field differences, and for this reason field-normalization may not play an important role.

At the third and fourth positions in the validity ranking of the indicators are CSNCR and $I3$ with a very similar value. Both indicators also differ scarcely from (perform slightly better than) RCR, MNCS, and the three SNSCI indicators (as well as the citation counts measured over the variable citation window until 2017). Thus, the newly developed $I3$ indicator holds up well against many other (field-normalized) indicators by discriminating equal to (or even slightly better than) the other indicators between the four F1000 quality classes.

With regard to percentiles (InCites and Hazen percentiles), our results are in disagreement to the previous results of Bornmann and Leydesdorff (2013). They reported very positive results for citation percentiles when this indicator is compared with other (field-normalized) indicators: "Percentile in Subject Area achieves the highest correlation with F1000 ratings" (p. 286). Using other data, the results in this study show, however, that percentiles (InCites and Hazen percentiles) perform comparably worse. The reasons for the differences between both studies should further be investigated in future studies.

A reason for the comparably poor performance of some of the percentile-based indicators might be that the F1000Prime data is a selective group of papers regarded as especially useful for other researchers in biomedicine. Therefore, a discrimination of these papers with respect to their quality scores focuses on a rather high level of quality (very high quality vs. high quality). This suggests that percentile-based indicators focusing on the upper end of the citation distribution (especially the top-1% indicator) are better suited for adequately discriminating this specific data, whereas indicators considering other parts of the distribution may have less discriminative power in this set.

Furthermore, even in the already selective F1000 dataset, highly skewed distributions of quality scores and indicator values can be observed. Most of the papers fall into F1000 classes 1 or 2 which are very similar when compared to indicators which include low quality scores. This is also reflected in the indicator values across classes: for most of the indicators, classes 1 and 2 are rather similar, whereas class 4 substantially differs from the other classes. As a result, the assessment of the indicators' validity mainly rests on the ability to discriminate the top papers from the rest of the (already selective set of) papers. This may also favor percentile-based indicators focusing on the upper end of the citation distribution. We expect that other percentile-based indicators would be better able to differentiate between papers (groups of papers) reflecting the broad range of different quality levels.

Although many (field-normalized) indicators which we included in this study might measure citation impact similarly, the results of our study also show that the concordance between the indicators is not perfect. The use of certain (field-normalized) indicators in research evaluation might lead to different results on citation impact – depending on the used indicators. Against the backdrop of our results concerning differences between the indicators, it might be interesting to investigate in future studies, whether there are particular papers or types of papers which differ significantly between the various indicators. Information like this would be valuable in pointing out the biases of the various indicators.

# Acknowledgements

The bibliometric data used in this paper are from an in-house database developed and maintained in cooperation with the Max Planck Digital Library (MPDL, Munich) and derived from the Science Citation Index Expanded (SCI-E), Social Sciences Citation Index (SSCI), Arts and Humanities Citation Index (AHCI) prepared by Clarivate Analytics, formerly the IP & Science business of Thomson Reuters (Philadelphia, Pennsylvania, USA).